\documentclass[apjl]{emulateapj}

\usepackage[usenames]{color}
\usepackage{mathrsfs}
\usepackage{graphicx}
\newcommand{\vvec}{{\vec{v}}}

\newcommand{\Bvec}{{\vec{B}}}
\newcommand{\Fvec}{{\vec{F}}}
\newcommand{\er}  {{\vec{e}_r^{}}}
\newcommand{\ez}  {{\vec{e}_z^{}}}

\newcommand{\epar} {{\vec{e}_\parallel^{}}}
\newcommand{\eperp} {{\vec{e}_\perp^{}}}
\newcommand{\ephi}{{\vec{e}_\varphi^{}}}
\newcommand{\muzero}{{\mu_0^{}}}
\newcommand{\fmlong}{{{F_M^{}}_{long}^{}}}
\newcommand{\grav}{{{g}}}

\newcommand{\machpolsq}{{M_{A_p}^2 }}

\DeclareGraphicsExtensions{.pdf,.png,.jpg}
\newcommand{\commentf}[1]{{\hskip -3pt}}

\shorttitle{Solar Tornadoes}
\shortauthors{Luna, Moreno-Insertis \& Priest}

\begin{document}

\title{Are tornado-like magnetic structures able to support solar prominence plasma?}

\author{M. Luna\altaffilmark{1,2}, F. Moreno-Insertis\altaffilmark{1,2} \& E. Priest\altaffilmark{3}}

\altaffiltext{1}{Instituto de Astrof\'{\i}sica de Canarias, E-38200 La Laguna, Tenerife, Spain}
\altaffiltext{2}{Departamento de Astrof\'{\i}sica, Universidad de La Laguna, E-38206 La Laguna, Tenerife, Spain}
\altaffiltext{3}{Mathematics Institute, University of St Andrews St Andrews KY16 9SS, UK}
\begin{abstract}
Recent high-resolution and high-cadence observations have
surprisingly suggested that prominence barbs exhibit apparent rotating
motions suggestive of a tornado-like structure. Additional evidence has been
provided by Doppler measurements. The observations reveal opposite velocities
for both hot and cool plasma on the two sides of a prominence barb. This
motion is persistent for several hours and has been interpreted in terms of
rotational motion of prominence feet. Several authors suggest that such barb
motions are rotating helical structures around a vertical axis similar to
tornadoes on Earth. One of the difficulties of such a proposal is how to
support cool prominence plasma in almost-vertical structures against
gravity. In this work we model analytically a tornado-like structure and try
to determine possible mechanisms to support the prominence plasma. We have
found that the Lorentz force can indeed support the barb plasma provided the
magnetic structure is sufficiently twisted and/or significant poloidal flows
are present. 
\end{abstract}

\section{Introduction}
Recent high-resolution and high-cadence observations have revealed a possible
rotating motion of prominence barbs around a vertical axis. These have been
called barb tornadoes \citep{priest2014} due to the similarity of the
apparent motion of such structures on the limb to terrestrial tornadoes.
\citet{su2012} reported an event where a prominence shows an apparent
rotating motion with velocities of up to $8 \, \mathrm{km \, s^{-1}}$. The
observation shows cool prominence plasma seen in absorption in the $171$
\AA\ passband of the SDO/AIA instrument. The authors argue that the motion is
a rotation projected on the solar limb plane. However, such projected motions
are also compatible with oscillations and counter-streaming flows, as pointed
out by \citet{panasenco2014}. More events, up to 201 barb-tornadoes, have
been reported by \citet{wedemeyer2013} using AIA 171\AA\ images. However, these authors studied mainly the morphology of the barbs and deduced a wide range of sizes and lifetimes.

\citet{orozco-suarez2012} reported Doppler shifts using \ion{He}{1}~1083.0~nm triplet data from the
German Vacuum Tower Telescope (VTT) of the Observatorio del Teide (Tenerife,
Spain). The observations revealed opposite velocities at the edges of the
prominence feet of $\pm \, 6 \, \mathrm{km \, s^{-1}}$ along a slit placed
close to the solar surface suggesting rotation of the cool prominence feet. The width of the feet is about $20\arcsec$ indicating a period of rotation of 4 hours. 

More recently, \citet{su2014} also reported Doppler shifts in a prominence
pillar using the \ion{Fe}{12}~195\AA\ line of the EIS instrument onboard the Hinode satellite. The observations revealed a bipolar
velocity pattern along the whole vertical prominence pillar. The velocity is
almost zero at the tornado axis and increases linearly up to $\sim \pm 5 \,
\mathrm{km \, s^{-1}}$ at the two edges of the observed structure. This
indicates that also the million-degree plasma related to the tornado-like
prominence may be rotating. The EUV bands of SDO/AIA reveal that the cool
plasma seen in absorption moves in consort with the hot plasma. Additionally, \citet{martinez-gonzalez2015}, have found evidence of helical magnetic structure simultaneously at two prominence feet. All this
evidence suggests that barb tornadoes are rotating vertical structures, nevertheless more observational evidence is needed.


The existence of such structures opens new questions concerning the origin of
the tornado rotation and the influence of the rotating barb on the rest of
the filament. \citet{wedemeyer-bohm2012} and \citet{su2012} proposed that the
barb-tornadoes are driven by photospheric vortex flows of the kind observed
by \citet{brandt1988} and \citet{bonet2008}: according to this view, the barb
field lines are rooted in the vortices and the latter's rotating motion is
transferred to the barb. The authors also proposed that the barb tornadoes
inject chromospheric plasma and helicity into the upper filament throughout
the rotating barb. However, such vortex flows have yet not been observed at all below barbs, let alone as a matter of course.

Another possibility for the origin of spiral motions is three-dimensional reconnection (at or above the photosphere) during cancellation of photospheric magnetic fragments, since such reconnection will naturally produce vortex motions \citep[e. g.,][]{priest2014} and could also fuel a prominence with mass.  A third possibility arises from the fact that a prominence represents a concentration of magnetic helicity in twisted magnetic fields. Thus, if part of a prominence dips down towards the photosphere, it is possible that such magnetic helicity and its associated twisting motions may be focused in the dip.

In this letter we discuss how the massive cool plasma is supported against gravity in a helical magnetic structure. We find that the barb-tornadoes bear many similarities to astrophysical plasma jets in which magneto-centrifugal forces accelerate the plasma. By using recent current tornado data and typical barb parameters we conclude that it is actually possible for the magnetic force to support and accelerate the cool barb plasma against gravity provided the structure is highly twisted and/or significant poloidal flows are present.

\section{The barb-tornado model}

\begin{figure}
\begin{center}
\includegraphics[width=0.5\textwidth]{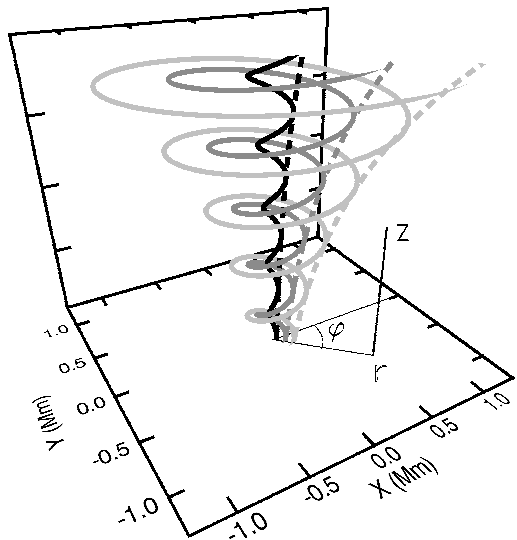}
\end{center}\caption{Schematic picture of a tornado-like magnetic structure using the elementary force-free field of \citet{schatzman1965}. Solid lines are the three dimensional representation of the magnetic field lines. Dashed lines are the poloidal field lines in the plane $\varphi = \mathrm{constant}$.}
\label{fig:cartoon}
\end{figure}

Inspired by the observations, we consider in the following an
axisymmetric model rotating around a vertical axis. We use cylindrical
coordinates $(r,\varphi,z)$ with $z$ coinciding with the rotation
axis. The axisymmetry condition implies that all quantities are
$\varphi$-independent: $\partial / \partial \varphi =0$. An important
restriction on the axisymmetric magnetic field is that the field lines should
become vertical as we approach the rotation axis ($r\to 0$). The observed
structure appears to be stationary with no important changes of shape in the
EUV images and Doppler pattern for several hours. The Alfv{\'e}n speed is
around one thousand $\mathrm{km\,s^{-1}}$ in the corona and of order one
hundred $\mathrm{km\,s^{-1}}$ in the cool prominence plasma.  Thus, the
travel time of a magnetic perturbation is less than a minute along the
vertical axis of the tornado. This indicates that, during the few hours the observation, the system has plenty of time to relax
and produce a stationary magnetic structure. We therefore set 
$\partial/\partial t=0$ in the equations. In this situation the MHD 
induction and momentum equations become, respectively, 
\begin{eqnarray}
0 &=&  \nabla \times (\vvec\times \Bvec)
~,\label{eq:mhd1} \\
0 &=&  - \rho (\vvec \cdot \nabla)
\vvec -\nabla p + \frac{1}{\muzero}(\nabla \times \Bvec)\times
\Bvec+\rho \vec{g}\, , \label{eq:mhd2} 
\end{eqnarray}
where $\rho$, $\vvec$, $\Bvec$, $p$ and $\vec{g}= - g \vec{e}_z$ are the plasma density, the velocity, magnetic field, gas pressure and gravity respectively.
Both the plasma velocity and magnetic field can be naturally decomposed into
\emph{poloidal} and \emph{toroidal} components, 
\begin{eqnarray}
\vvec &=& \vvec_p + v_\varphi\,\ephi ~,\\
\Bvec &=& \Bvec_p + B_\varphi\,\ephi ~,
\end{eqnarray}
with $\vvec_p\cdot\ephi=\Bvec_p\cdot\ephi = 0$. Only for illustration purposes we are showing in Figure \ref{fig:cartoon} the force-free solutions of \citet{schatzman1965}. The poloidal planes $(r,z)$ correspond to
$\varphi=\mathrm{constant}$ surfaces. Given the axisymmetry, we can easily
define the angular velocity $\Omega(r,z)$ using $\vec{v}_\varphi = r\,
\Omega(r,z)\,\ephi$. The advection (or inertial) term $ (\vvec \cdot \nabla)
\vvec$ can be written as 
\begin{equation}
 (\vvec \cdot \nabla) \vvec=(\vec{v}_p \cdot \nabla) \vec{v}_{p} -r\, \Omega^2\, \vec{e}_r + \frac{\vec{v}_p}{r} \cdot \nabla (r^2 \Omega) \vec{e}_\varphi ~,
\label{eq:inertial}
\end{equation}
i.e., advection in the poloidal plane plus two acceleration terms
associated with the prescribed rotation profile
$\Omega(r,z)$, the first of which is clearly the centripetal
force of a circular motion. 

In a stationary regime there must be force balance including the inertial
terms. For ease of notation, we use the symbol $\Fvec_{nm}$ for the
sum of pressure gradient, gravity and centrifugal force:
\begin{equation}\label{eq:nm_forces}
\Fvec_{nm}^{}\; {\buildrel \scriptstyle def \over =}\;  - \nabla p \;-\;\rho \grav
\ez\; + \; 
\rho r \Omega^2\er\; .
\end{equation}
The poloidal part of the equation of motion (\ref{eq:mhd2}) can then be split
into into its components parallel and perpendicular to $\Bvec_p$,
\begin{eqnarray}
\frac{\rho}{2} \frac{\partial \, v_p^2}{\partial s} &\;=\;&{\Fvec_{nm}^{}} \cdot \epar-\frac{1}{2 \muzero  r^2}
\frac{\partial(r B_\varphi)^2}{\partial s} ~, \label{eq:longitudinal_equilibrium_withflows}
\\ 
\noalign{\vspace{4mm}}
0 \; &\;=\;&\left[\Fvec_{nm}^{} - \nabla
  \frac{B_p^2}{2\,\muzero}-\frac{1}{2\,\mu_0\, r^2}\nabla(r
    B_\varphi)^2\right] 
 \cdot \eperp   \nonumber\\
&\;+\;& \left( 
\frac{B_p^2}{\muzero} -\rho \, v_p^2 \right) \kappa_{pol} 
\;, 
\label{eq:transverse_equilibrium_withflows}\\\nonumber
\end{eqnarray}
where $\epar$, $\eperp$ are the unit vectors in the poloidal plane parallel
and perpendicular to $\Bvec_p$, respectively, and $\kappa_{pol}$ and $s$ are
the curvature and arc-length along the poloidal field lines (the latter
is chosen such that $\epar$ points in the sense of growing $s$ illustrated as dashed lines in Fig. \ref{fig:cartoon}). For future
reference, we also note that $v_p$ and $B_p$ are defined through 
$\vvec_p \cdot \epar$ and $\Bvec_p\cdot \epar$, respectively, i.e., they can
be positive or negative. 
Equation~(\ref{eq:longitudinal_equilibrium_withflows}) represents the
steady flow of plasma along field lines and provides clues concerning
the support of the cool barb plasma; Equation
(\ref{eq:transverse_equilibrium_withflows}) represents the global transverse
equilibrium of the magnetic structure. The remaining component of the
equation of motion, the azimuthal component, is
\begin{equation}
\rho\,(\vvec_p \cdot \nabla)(r^2\,\Omega)  \;= \;
\left(\frac{\Bvec_p}{\mu_0^{}}\cdot\nabla\right) (r\,B_\phi) \;.\label{eq:motion_azimuthal}\\
\end{equation}

As an additional remark concerning Figure \ref{fig:cartoon}, note that in a barb-tornado structure the density along the poloidal field lines should be larger close to the axis than in the outer regions of the structure. The observation of \citet{su2014}, which shows a dense column in absorption tapering off with height, may well be compatible with a helically opening structure as shown in the figure.

\subsection{Pure Rotation}\label{sec:steady_rotation}

In a situation of pure rotation there is no plasma flow along the poloidal
field ($\vvec_p =0$), and the inertial term (\ref{eq:inertial}) reduces to the
centripetal component $\,-\rho \, r \,\Omega^2\,\er\,$.
In this case, from Equation~(\ref{eq:motion_azimuthal}), we see that the
azimuthal component of the Lorentz force is zero, i.e., $r B_\varphi$ is
constant along each poloidal field line. On the other hand, using the induction
equation one can easily see that in this case $\Omega$ is constant along each given
field line, in agreement with Ferraro's
isorotation law \citep{ferraro1937}. The magnitudes of $r B_\varphi$ and
$\Omega$ are then determined by the boundary conditions of the problem. Given
the constancy of $r B_\varphi$, we see from 
the longitudinal equilibrium equation
(\ref{eq:longitudinal_equilibrium_withflows}) that the Lorentz force has no
longitudinal component, and so the equation reduces to
\begin{equation}\label{eq:longitudinal_equilibrium}
{\Fvec_{nm}^{}} \cdot \epar = 0 \; .
\end{equation}
The Lorentz force is thus purely poloidal and perpendicular to $\Bvec_p$. Calling $\theta$ the angle between the poloidal field and the horizontal
direction, equation (\ref{eq:longitudinal_equilibrium}) may be written
\begin{eqnarray}\label{eq:eq-decomposed}
r\,\Omega^2 \,\cos \theta -\frac{1}{\rho}\,\frac{\partial p }{\partial s}
-\,\grav \, \sin \theta =0\;,
\end{eqnarray}
implying a balance of the non-magnetic forces. 
Note, in particular, that the centrifugal term could  help support the plasma against
gravity if the field lines are sufficiently close to horizontal (i.e.,
$\sin\theta > 0$ sufficiently small).

Is the foregoing purely-rotating stationary equilibrium a realistic
possibility for the observed apparent barb tornadoes? The latest observations
\citep{su2014} reveal a rotational velocity of $5 ~\mathrm{km~s^{-1}}$ at the
edges of the structure with $r=2\arcsec=1.5 ~\mathrm{Mm}$, so 
 $\Omega
\approx 3 \times 10^{-3} ~\mathrm{rad ~s^{-1}}$. \citet{wedemeyer2013} found average values of barb widths of $r\approx2\arcsec$ in agreement with \citeauthor{su2014}. With these values we can
compare the centrifugal acceleration to the gravitational acceleration,
\begin{equation}\label{eq:centrifugal_force_ratio}
\frac{r\,\Omega^2}{\grav} \approx 0.06 = \tan(3.4\degr)\;.
\end{equation}
With the values found by \citet{orozco-suarez2012} this ratio is even smaller. The only way to have centrifugal support of the plasma is then for the field
lines to be almost horizontal, which contradicts the tornado picture. Can a
pressure gradient help support the plasma in a non-horizontal field?  Hot,
coronal plasma, can be supported by a pressure gradient against gravity
across coronal distances even in vertical field lines. However, for cool
prominence plasma the pressure scale-height is too small to balance gravity
in barbs as tall as those observed. Even if the field lines close to the
rotation axis were filled with hot coronal plasma, the magnetic structure
would have to turn almost horizontal (say, $3 \degr$, as given in
Eq.~\ref{eq:centrifugal_force_ratio}) in the region holding cool prominence
plasma. We conclude that the centrifugal force associated with the rotation
cannot support the cool barb plasma against gravity in a helical field
structure.

Another way to illustrate this conclusion is to estimate the rotational
speed necessary to have purely centrifugal support, namely
$v_\varphi = \sqrt{r \grav \tan \theta}\;.$
Assuming a magnetic field inclination of $45\degr$, say, at the edges of the
observed barb tornadoes, the rotation velocity would have to be
$20~\mathrm{km~s^{-1}}$ if we use the data of \citet{su2014} and
$45~\mathrm{km~s^{-1}}$ when using those of \citet{orozco-suarez2012}, much
larger than the measured speeds.

\subsection{General case with poloidal flows}

We consider now a more general scenario allowing for flows in the poloidal
direction. Now the simple situation of constant $\Omega$ and $r B_\varphi$
along the field lines no longer applies. Checking for instance the azimuthal
equation of motion (\ref{eq:motion_azimuthal}), we see that a change of
specific axial angular momentum of the plasma elements associated with the
poloidal motion must be associated with a non-zero toroidal component of the Lorentz force,
$(\Bvec_p/r)\cdot\nabla(r\,B_\varphi) \, \ephi$. Hence, in general $r
B_\varphi$ can no longer be constant along field lines. 
In this situation, the Lorentz force also has a non-zero projection along poloidal field
lines (Eq. \ref{eq:longitudinal_equilibrium_withflows}), $\fmlong$,
\begin{eqnarray}\label{eq:magforcegeneral}
\fmlong & {\buildrel \hbox{\tiny def} \over = } 
& -\frac{1}{2 \muzero r^2}\frac{\partial(r B_\varphi)^2}{\partial
  s} \;.
\end{eqnarray}

In spite of the added complication of this general scenario, there is a set
of conserved quantities along the field lines. One can obtain them by
following the general procedure used in the theory of astrophysical jets
\citep{mestel1961,lovelace1986}. The induction equation requires that
$\nabla\times(\vvec_p \times \Bvec_p) =0 $, so the poloidal velocity must 
be parallel to the poloidal field lines, for otherwise a singularity would
arise in the toroidal electric field at the $z$-axis. So, we can write 
\begin{equation}
\vec{v}_p= \kappa(r,z) ~\vec{B}_p \;.\label{eq:poloidal_flow_parallel_to_field}
\end{equation}
Integrating the MHD equations along field lines and simplifying the
resulting expressions, a set of conserved quantities results, namely, 
\begin{eqnarray} \label{eq:cons1}
\muzero \rho \kappa &=& K~, \\ \label{eq:cons2}
\Omega - \frac{K B_\varphi}{\muzero \rho \,r}    &=& W~, \\ \label{eq:cons3}
\Omega r^2 -\frac{r B_\varphi}{K} &=& \Lambda~,
\end{eqnarray}
where $K$, $W$, and $\Lambda$ are all constant along each poloidal field line. 
Those relations allow us to find an explicit expression for $r B_\varphi$ along the poloidal field.  One can write it in terms of the {\it poloidal
  Alfv\'en Mach number}  
\begin{equation}\label{eq:machpol_and_rho}
\machpolsq {\buildrel \hbox{\tiny def}\over = }\frac{v_p^2}{B_p^2/\muzero
  \rho}=\frac{K^2}{\muzero \rho} ~,
\end{equation}
and of the {\it Alfv\'en radius}, $r_A$,  defined by
\begin{equation}\label{eq:alfen_radius}
r_A^2 \;{\buildrel \hbox{\tiny def}\over =} \; \frac{\Lambda}{W} \;,
\end{equation}
as follows:
\begin{eqnarray} 
r B_\varphi&=&K W \frac{ r^2- r_A^2}{1-\machpolsq} \label{eq:rbphik}\;. 
\end{eqnarray}
Expression (\ref{eq:rbphik}) become singular when
$\machpolsq \to 1$.  In the classical wind solutions, the flow speed increases from
sub-Alfvenic to super-Alfvenic at a given point, and, to avoid the singularity,
this transition must happen precisely at the Alfv\'en radius, $r=r_A$. In
such models, this requirement serves as an internal boundary condition to
pick up the desired trans-Alfvenic solution instead of the a completely sub-Alfvenic or
completely super-Alfvenic solution. In our case, however, it is unlikely that the flows
reach Alfv\'enic values within the tornado; rather, as shown below, we expect
$\machpolsq$ to remain below $1$ throughout.

Following Eq.~(\ref{eq:longitudinal_equilibrium_withflows}) and the results
of Section~\ref{sec:steady_rotation}, to support the cool plasma along
field lines the magnetic force $\fmlong$ of Eq.~(\ref{eq:magforcegeneral})
should be larger than the centrifugal force, $f_C = \rho\, \Omega^2 \, r$.
From Eqs.~(\ref{eq:cons1}) and (\ref{eq:cons3}), the ratio of these forces is
\begin{eqnarray}\label{eq:comparison-mag-cent}
\left| \frac{\fmlong}{f_C} \right|= \left| \frac{T_B}{T_v} \right| \left( \frac{r}{L_{\Omega}} \right) \;,
\end{eqnarray}
in terms of the ratios ($T_B$ and $T_v$) of azimuthal to
poloidal components for the magnetic fields and plasma velocities:
\begin{equation}
T_B {\buildrel \scriptstyle def \over =}
  \frac{B_\varphi}{B_p} \quad\hbox{and}\quad
T_v {\buildrel \scriptstyle def \over =}   \frac{v_\varphi}{v_p}\;,
\end{equation}
and the length $L_{\Omega}$, defined as
\begin{equation}
L_{\Omega} = \left| \frac{\partial \ln (r^2 \Omega)}{\partial s}
\right|^{-1}\;.\label{eq:charac_length_omega} 
\end{equation}
The characteristic length ($\ref{eq:charac_length_omega}$) can be assumed to be
of order the radial size of the system, i.e., $L_{\Omega} \sim \hbox{O(r)}$. Hence, from Equation
(\ref{eq:comparison-mag-cent}), it is necessary for the value of the parameter $\left|T_B/T_v \right|$ to be large in order to have a
barb-tornado dominated by the magnetic force. We see that a high level of
magnetic twist, and a comparatively large poloidal velocity (compared to the
rotational velocity) are necessary to provide magnetic support. 

\begin{figure}
\includegraphics[width=8cm]{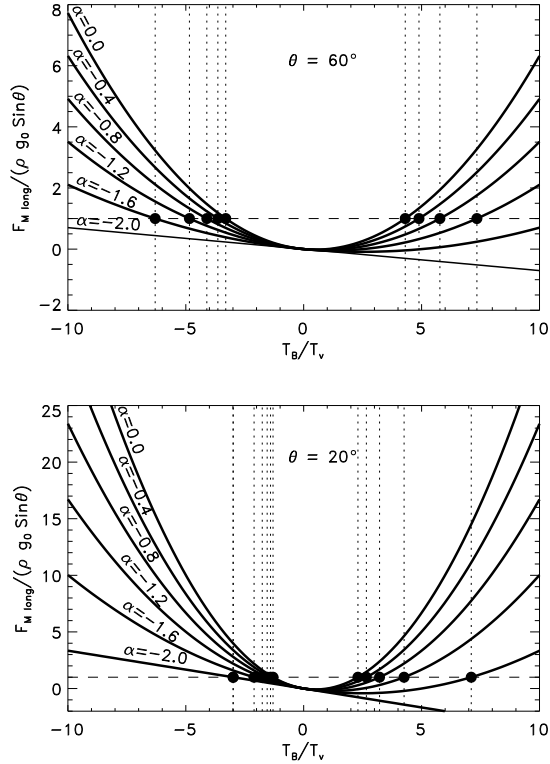}
\caption{Magnetic force along the field lines compared with the longitudinal component of gravity. For magnetic support, the value $1$ must be reached (marked by dots for the different curves)}
\label{fig:ratio}
\end{figure}

We can derive a more exact estimate of the possibilities of magnetic support
of tornado plasma against gravity using expression
(\ref{eq:magforcegeneral}) 
for the longitudinal magnetic force combined with expression
(\ref{eq:rbphik}). First of all, we can use the definitions 
of $W$ and $K$ to write their product in the form 
\begin{equation}
K\,W = \machpolsq \frac{B_p}{r} \left(T_v -
  T_B\right)
\end{equation}
keeping in mind, nevertheless, that $K$, $W$ (and $r_A$) are constant along
each field line. 
We can now derive Eq.~(\ref{eq:rbphik}) with respect to the arc-length and, after a little algebra,  obtain the general expression 
\begin{equation}\label{eq:fmlong_estimate_1}
\fmlong\;=\; \frac{2}{r} \frac{B_\varphi\,B_p}{\mu_0}
\frac{\machpolsq}{1-\machpolsq} \cos \theta \left[\,\left(1+\frac{\alpha}{2}
  \right) \, T_B-T_v \right] \, 
\end{equation}
where we have used $\cos \theta = \partial r / \partial s$, and called $\alpha$ the scale
of variation of $\rho$ along poloidal field lines in terms of the
cylindrical radius, as follows:
\begin{equation}
\alpha\quad {\buildrel \scriptstyle def \over =} \;
\frac{r}{\cos \theta} \frac{\partial \log \rho}{\partial s}\;. 
\end{equation}
For the estimate that follows, it is best to write Equation~(\ref{eq:fmlong_estimate_1}) in terms of an observed quantity,
namely $v_\varphi$, and of the ratio $T_B/T_v$, which turns out to be the essential dimensionless variable in the resulting expression. Further, to
calibrate the possibilities of magnetic support, we normalize (\ref{eq:fmlong_estimate_1}) with respect to gravity along
the poloidal field, $\rho \, \grav\,\sin\theta$. The result is:
\begin{eqnarray}\nonumber
\frac{\fmlong}{\rho\,\grav\,\sin\theta} \quad&=&\quad \frac{2}{1-\machpolsq}\; 
\left(\frac{v_\varphi^2}{\grav\,r}\right) \;
\,\cdot\; \\
\noalign{\vspace{3mm}}
&& \left(\frac{T_B}{T_v}\right)  \left[
\, \left(1+\frac{\alpha}{2} \right)\, \frac{T_B}{T_v} -1 \right] \cot\theta\, .
\label{eq:support}
\end{eqnarray}
For the applications to prominence barbs we envisage cases with $\machpolsq
<1 $. The term $1-\machpolsq$ in Eq.~\ref{eq:support} therefore favors
magnetic support. In fact, in many cases $\machpolsq \ll 1 $ and so for simplicity this is the case we consider. Assuming the value of the ratio
$v_\varphi^2/(\grav\,r)$ to be given from observations expression (\ref{eq:support}) is basically a quadratic polynomial in $T_B/T_v$
with parameters $\alpha$ and $\theta$. In figure~\ref{fig:ratio}, we show (\ref{eq:support}) for $\theta=60 \degr$ (upper panel) and $\theta=20
\degr$ (lower panel), for different relevant values of $\alpha$. To draw the
figure we have used $v_\varphi^2/(\grav\,r) = 0.06$ (i.e., $v_\varphi= 5$ km
s$^{-1}$ and $r = 1.5$ Mm), which is the value that led us in
Sec.~\ref{sec:steady_rotation}, Equation~\ref{eq:centrifugal_force_ratio}, to
conclude that there is no possible support for the plasma in the purely
rotating case. 
For magnetic support, the relevant stretches of the curves are those near the
dashed horizontal line at ordinate $= 1$. For ease, we have marked the
cut of each curve with that horizontal line with a large black dot and a
thin dotted vertical line. The following guidelines have been used in the
choice of values for $\alpha$: in a solar tornado $\alpha$ is
probably negative, i.e., the density decreases as one goes outward along the
tornado field lines. Also, $|\alpha|$ should not be larger than
of order unity, since the lengthscale of variation of $\rho$ should be not too
different from $r$ itself. Finally, in the figure we restrict ourselves to
concave-upward parabolas (i.e, $\alpha > - 2$), since those are the most
favorable cases for magnetic support. 

We see that it is not difficult to find values of $T_B/T_v$ that yield
magnetic support of the plasma against gravity. In all cases shown, though,
the basic variable $|T_B/T_v|$ must be above $1$: for $\theta
= 60 \degr$, for instance, the values marked in the figure range from
$-6.3$ to $-3.3$, on the negative side and from $4.3$
to $7.3$ on the positive side. Note that the actual ranges are $[-14.2,-3.3]$ and $[4.3, \infty ]$. For $\theta=20\degr$, the absolute values are
smaller. We expect $T_v$ to be perhaps $1$ or $2$, reflecting the fact that
the outward motion of the tornado is possibly as fast as 
the rotating motion or a little less so. Thus, for magnetic support one would
need a substantial level of magnetic twist, possibly $|B_\varphi / B_p|$ from a few to several
units. If the poloidal flow becomes less important, then the amount of
magnetic twist would increase to unrealistic values. 

%

\section{Conclusions}
In this work we have investigated a possible mechanism to support dense plasma in a prominence barb against gravity. We have modeled a barb tornado as an axisymmetric structure with rotation about the symmetry axis, in which the magnetic field is vertical close to the tornado axis. Pressure gradients are ruled out as a support for the plasma because the small pressure scale height of prominence plasma implies that they could support vertical structures only a few hundred kilometers tall, much smaller than the height of observed barbs.  

In a barb tornado with a rotating helical field, extra magneto-centrifugal forces are present. We have found that the centrifugal force is much smaller than solar gravity for the barb-tornadoes observed so far. However, the poloidal magnetic force is a good candidate to support cool prominence plasma or even to inject such plasma into a prominence. For that, the structure must have significant magnetic twist and/or poloidal flows not much smaller than the rotational velocities. Whether this is or otherwise the case is a question for future observations. However, more theoretical work and observational evidence are needed to elucidate the origin of poloidal flows or magnetic twist.


\acknowledgments
M. Luna and F. Moreno-Insertis acknowledge the support by the Spanish Ministry of Economy and Competitiveness through projects AYA2011-24808 and AYA2014-55078-P. M. L. is also grateful to ERC-2011-StG 277829-SPIA. E. R. Priest is grateful to the UK STFC and the Leverhulme Trust for financial support. The authors also are grateful to D. Orozco-Su\'arez and K. Knizhnik for helpful discussions.


\end{document}